\documentclass[preprint,12pt]{elsarticle}
\usepackage{amsmath, amsthm, amssymb}

\begin{document}
\title{On the characteristic polynomial of an effective Hamiltonian}
\author{Yong Zheng  \corref{correspondingauthor}}
\cortext[correspondingauthor]{Corresponding author}
\ead{zhengyongsc@sina.com}
\address{School of Physics and Electronics, Qiannan Normal University for Nationalities,
	Duyun 558000, China}

\begin{abstract}
The characteristic polynomial of the effective Hamiltonian for a
general model has been discussed. It is
found that, compared with the associated energy eigenvalues,   this characteristic polynomial generally has better analytical properties and
larger convergence radius when  being expanded in powers of the
interaction parameter, and hence is more suitable for a perturbation
calculation. A form of effective Hamiltonian which has the same
singularities (branch points) as such  characteristic polynomial has
also been  constructed.
\end{abstract}


\maketitle

\section{Introduction}

Many quantum interaction models, such as the Hubbard model and nuclear-shell model, involve states with different energy scales. If one is only interested in some special states, say, the lower-energy-scale states, the full Hilbert space of the model can be partitioned into two disjoint subspaces: the so-called $ P $- and $ Q $-spaces, which are spanned by the states we are interested and uninterested in respectively. For a full Hamiltonian $H(\lambda)$ with an interaction parameter $ \lambda $, an effective elimination of the states we are uninterested in, via rigorous perturbation treatment or canonical transformation, etc.,  can yield a dimension-reduced equivalent Hamiltonian  which only acts on the $ P $-space or the remaining states, the so-called effective Hamiltonian $ H_{\mathrm{eff}}(\lambda) $ (For recent and historical references of this, see Refs.~\cite{HS1, HS2, HS3, HS4, HS5, CT1,CT2}).

The obtained $ H_{\mathrm{eff}}(\lambda) $ generally is complicated and always in an infinite series form of the interaction parameter $\lambda$, from which the further solving process for interesting physical quantities such as the energy eigenvalues is still nontrivial. On the other hand, the convergence problem of such series-form  of $ H_{\mathrm{eff}}(\lambda) $ itself is difficult to discuss \cite{HS1, HS2}. Especially, the effective Hamiltonian can be  derived by several ways, e.g., by block-diagonalization of $H(\lambda)$ via similarity transformation, Rayleigh-Schr\"{o}dinger or Brillouin-Wigner Perturbation.  Different derivation procedures usually lead to  different series-forms of $ H_{\mathrm{eff}}(\lambda) $, which can be energy-dependent or energy-independent, Hermitian or Non-Hermitian \cite{HS4, HS5}.  Mathematically, even for the same problem, there are infinitely many $ H_{\mathrm{eff}}(\lambda) $ which are similar matrices to each other \cite{BL1, BL}. This makes the theory of effective Hamiltonian lack unity in some sense.

It is interesting whether the  effective Hamiltonian can be discussed in a more unified and efficient way. Here we want to study $ H_{\mathrm{eff}}(\lambda) $ from another point of view, i.e., via discussing its characteristic polynomial $\det[E-H_{\mathrm{eff}}(\lambda)]$. One consideration of our discussion is that though the effective Hamiltonian can be in different Matrix forms similar to each other,  the characteristic polynomial is unique. This may enable us to study some essential properties of the effective Hamiltonian itself, rather than those only associated with its some particular form. Additionally, we will find that the characteristic polynomial itself has some very important characteristics,  such as that it always has less singularities in the
complex plane of $ \lambda $ than the $ P $-space energy eigenvalues,  which would allow us to perform a more effective perturbation treatment even when the detailed form of $ H_{\mathrm{eff}}(\lambda) $ is unknown.

\section{Formulation}

We consider a system of dimension $ \tilde{N}=N+M $, of which the model space can be decoupled into two subspaces: a $P$-space spanned by $N$ states of interesting and a $Q$-space spanned by the other $M$ states. The Hamiltonian can be written as
\begin{equation}\label{H}
	H(\lambda)=H_0+\lambda H_I,
\end{equation}
where $ H_0 $ and $ H_I $ both are Hermitian:
\begin{align*}
	&H_0=\sum_{n=1}^{N}\epsilon_{n}^{_P}|\psi^{_P}_n\rangle \langle \psi^{_P}_n|+\sum_{n=1}^{M}\epsilon_{n}^{_Q}|\psi^{_Q}_n\rangle \langle \psi^{_Q}_n|,\\
	&H_I=\sum_{n,m} h^{\!_P}_{mn}|\psi^{_P}_n\rangle \langle \psi^{_P}_m|+\sum_{n,m} h^{\!_Q}_{mn}|\psi^{_Q}_n\rangle \langle \psi^{_Q}_m| +\sum_{n,m} \left[ h^{\!_{P\!Q}}_{mn}|\psi^{_P}_n\rangle \langle \psi^{_Q}_m|+\text{h.c}\right] ,
\end{align*}

Without losing generality, we assume that the unperturbed energies of $P$- and $Q$-space states all are non-degenerate, i.e.,  $ \epsilon_1^{_P} \neq \epsilon_2^{_P}$, $ \epsilon_1^{_P} \neq \epsilon_1^{_Q}$, etc. For small $ \lambda $, the eigenvalues of $P$-space states, $ E_1^{_P}(\lambda ), E_2^{_P}(\lambda ),\cdots, E_{N}^{_P}(\lambda )$ can be expanded as
\begin{subequations}
	\begin{align}
		&E_1^{_P}=\epsilon_1^{_P}+\lambda h^{\!_P}_{11}+ \lambda^2\left( \sum_{n\neq
			1}\frac{|h^{\!_P}_{1n}|^2}{\epsilon_1^{_P}-\epsilon_n^{_P}}
		+ \sum_{n}\frac{|h^{\!_{P\!Q}}_{1n}|^2}{\epsilon_1^{_P}-\epsilon_n^{_Q}}\right) +\cdots \tag{2.1} \label{EP1}\\
		&E_2^{_P}=\epsilon_2^{_P}+\lambda h^{\!_P}_{22} + \lambda^2\left( \sum_{n\neq 2}\frac{|h^{\!_P}_{2n}|^2}{\epsilon_2^{_P}-\epsilon_n^{_P}}
		+ \sum_{n}\frac{|h^{\!_{P\!Q}}_{2n}|^2}{\epsilon_2^{_P}-\epsilon_n^{_Q}}\right) +\cdots  \tag{2.2} \label{EP2}\\
		&\quad \cdots \cdots \notag\\
		&E_N^{_P}=\epsilon_N^{_P}+\lambda h^{\!_P}_{N\!N}+ \lambda^2\left(
		\sum_{n\neq
			N}\frac{|h^{\!_P}_{Nn}|^2}{\epsilon_N^{_P}-\epsilon_n^{_P}} +
		\sum_{n}\frac{|h^{\!_{P\!Q}}_{Nn}|^2}{\epsilon_N^{_P}-\epsilon_n^{_Q}}\right)
		+\cdots.  \tag{2.N} \label{EPN}
	\end{align}
\end{subequations}
Similar expressions can also be written out for the  eigenvalues of
$Q$-space states $ E_1^{_Q}(\lambda ), E_2^{_Q}(\lambda ),\cdots,
E_{M}^{_Q}(\lambda )$. These expansion series always break
down when $ \lambda $ is adequately large.  This is due to the fact
that the $ E_n^{_P}(\lambda ) $ and $ E_m^{_Q}(\lambda ) $, as
functions of $ \lambda $,  generally have singularities in the
complex plane of $ \lambda $. The convergence radius of the
expansion series is determined by the module of the singularity
closest to the origin, for each $ E_n^{_P}(\lambda ) $ or $ E_m^{_Q}(\lambda ) $.
Mathematically, since $ E_n^{_P} $ and $ E_m^{_Q} $
are solutions of the algebraic equation $\det[E-H(\lambda)]=0$, the only singularities are branch
points \cite{Ab1,Ab2, Ab3}.
Here,
\begin{multline}\label{det}
	\det[E-H(\lambda)]=\prod_{n=1}^{N}[E- E_n^{_P}(\lambda )]\prod_{m=1}^{M}[E- E_m^{_Q}(\lambda )]\\
	=E^{\tilde{N}}+p_1(\lambda)E^{\tilde{N}-1}+\cdots+p_{\tilde{N}-1}(\lambda)E+p_{\tilde{N}}(\lambda),
\end{multline}
with the coefficients $p_n(\lambda)$ being polynomials of $ \lambda $.

As is well-known, these branch points are characterized by ``level-crossing'', where two (or more) eigenvalues, accompanying their eigenfunctions, coincide.
Level-crossings can occur between two $P$-space states ($PP$-crossing), two $Q$-space states ($QQ$-crossing), or one $P$-space state  and one $Q$-space state  ($PQ$-crossing). Hence, if some $ \lambda $ in the
complex plane is a branch point of eigenvalue $E_n^{_{P(Q)}}(\lambda )$, it must also be a branch point of at least one other eigenvalue. For each $E_n^{_P}(\lambda )$ or $E_m^{_Q}(\lambda )$, the branch-point set can be denoted by $\{\lambda^{_P}_{n,l_n},\lambda^{_{PQ}}_{n,l'_n}\}$ or $\{\lambda^{_Q}_{m,l_m},\lambda^{_{QP}}_{m,l'_m}\}$ respectively, with the $\lambda^{_{P(Q)}}_{i,l_i}$ or $ \lambda^{_{PQ(QP)}}_{i,l'_i} $ representing the branch points  due to the level-crossing of $E_i^{_{P(Q)}}(\lambda )$ with the other $E_j^{_{P(Q)}}(\lambda )$ or $E_j^{_{Q(P)}}(\lambda )$ respectively, which are numbered by $l_i(l'_i)=1,2,\cdots$. For example, if $E_1^{_{P}}(\lambda )$ only has five branch points, three due to $PP$-crossing and two due to $PQ$-crossing, we have $\{\lambda^{_P}_{1,l_1},\lambda^{_{PQ}}_{1,l'_1}\}=\{\lambda^{_P}_{1,1}, \lambda^{_P}_{1,2},\lambda^{_P}_{1,3},\lambda^{_{PQ}}_{1,1},\lambda^{_{PQ}}_{1,2}\}$.

Then, the convergence radius of the $ E_n^{_P} $-series is determined by $r_n^{_P}=\min_{l_n,l'_n}\{|\lambda^{_P}_{n,l_n}|, |\lambda^{_{PQ}}_{n,l'_n}|\}$. We can further introduce a common convergence radius for these $ E_n^{_P} $-series,
\begin{equation}\label{rpb}
	\bar{r}^{_P}=\min_{n}\{r_n^{_P}\}=\min_{l_1, l'_1, \cdots,l_N, l'_N}\{\lambda^{_P}_{1,l_1},\lambda^{_{PQ}}_{1,l'_1},\cdots, \lambda^{_P}_{N,l_N},\lambda^{_{PQ}}_{N,l'_N}\}.
\end{equation}
Obviously,  $\bar{r}^{_P}$ specifies the maximum $|\lambda|$ for which all the $ E_n^{_P} $-series converge.

Mathematically, if some branch points, say these due to
$PP$-crossing, can  be removed, enlarged convergence radii may be
obtained for the perturbation calculation of $ E_n^{_P} $. Our
strategy is to use the characteristic polynomial of
$H_{\mathrm{eff}}(\lambda)$,
\begin{multline}\label{edetN}
	\det[E-H_{\mathrm{eff}}(\lambda)]=\prod_{n=1}^{N}[E- E_n^{_P}(\lambda )] =E^{N}-P_1(\lambda)E^{N-1}\\
	+\cdots+(-1)^{N-1}P_{N-1}(\lambda)E+(-1)^{N}P_{N}(\lambda) ,
\end{multline}
where the coefficients are symmetric polynomials of $P$-space eigenvalues,
\begin{subequations}
	\begin{align}
		&P_1(\lambda)=E_1^{_P}(\lambda )+E_2^{_P}(\lambda )+\cdots+E_N^{_P}(\lambda ), \tag{6.1} \label{P1}\\
		&P_2(\lambda)=\sum_{1\leq j_1<j_2\leq N}E_{j_1}^{_P}(\lambda )E_{j_2}^{_P}(\lambda ),\tag{6.2} \label{P2}\\
		& \quad \cdots \cdots  \notag\\
		&P\!_{N}(\lambda)=E_1^{_P}(\lambda )E_2^{_P}(\lambda )\cdots E_{N}^{_P}(\lambda
		). \tag{6.N} \label{PN}
	\end{align}
\end{subequations}

Unlike what in Eq.~\eqref{det}, since the detailed form of
$H_{\mathrm{eff}}(\lambda)$ has not been specified, the coefficients
$ P_n(\lambda) $ in Eq.~\eqref{edetN} is unknown; however,  we can
calculate them using Eqs.~\eqref{P1}--\eqref{PN}, i.e., by their relationship
with $ E_n^{_P}(\lambda ) $, since $ E_n^{_P}(\lambda ) $ can be
calculated
via the perturbation expansion as shown above. We first discuss the availability of such calculation for $ P_n(\lambda) $.

Noting that both $\det[E-H_{\mathrm{eff}}(\lambda)]$ and $ P_n(\lambda) $ are unchanged under the exchange of any two $P$-space eigenvalues, say, $ E_i^{_P}(\lambda ) $ and $ E_j^{_P}(\lambda ) $, one can expect that the  $PP$-crossing of $P$-space eigenvalues would not cause branch points for them. This can be demonstrated in more detail as follows. For any finite $\lambda$-region we want to discuss, Eq.~\eqref{det} can be further written  as
\begin{equation}\label{Ndet}
	\prod_{n=1}^{N}[E- E_n^{_P}(\lambda )]=\frac{\det[E-H(\lambda)]}{\prod_{m=1}^{M}[E- E_m^{_Q}(\lambda )]},
\end{equation}
where $E$ is viewed as a constant and takes its value large enough, say, outside the Gerschgorin disks of the $ E_m^{_Q}(\lambda ) $,   to ensure a nonzero denominator of the right-hand side of the equation. One should note that the left-hand side of this equation just has the form of $\det[E-H_{\mathrm{eff}}(\lambda)]$.

Although  $E_n^{_P}(\lambda )$ and $E_m^{_Q}(\lambda )$ each may have several branch points in the complex plane of $ \lambda $, the true branch points $\lambda_c$ of the right or left side of Eq.~\eqref{Ndet} can only be the those which are the common branch points  of the both sides.  Noting that here $\det[E-H(\lambda)]$, as shown in Eq.~\eqref{det}, has no branch point in the complex plane of $ \lambda $,  the only common branch points certainly are these due to the $PQ$-crossing of $E_n^{_P}(\lambda )$ and $E_m^{_Q}(\lambda )$, which constitute a set $S_{PQ}\equiv  \{\lambda^{_{PQ}}_{1,l_1}, \lambda^{_{PQ}}_{2,l_2},\cdots, \lambda^{_{PQ}}_{N,l_N} \}$, with the serial numbers $l_n=1,2,\cdots$ as above. Namely, unlike in the case of $E_n^{_P}(\lambda )$ and $E_m^{_Q}(\lambda )$,  the $PP$- or $QQ$-crossing does not cause any branch points for $\prod_{n=1}^{N}[E- E_n^{_P}(\lambda )]$ (i.e., $\det[E-H_{\mathrm{eff}}(\lambda)]$) and $ \prod_{m=1}^{M}[E- E_m^{_Q}(\lambda )] $.

Hence, the characteristic polynomial $\det[E-H_{\mathrm{eff}}(\lambda)]$, as a function of $\lambda$, is ``more analytical'' than $E_n^{_P}(\lambda )$, and  more suitable for a perturbation calculation.

To illustrate this more simply, we can first discuss the case of $N=2$:
\begin{equation}\label{2det}
	\det[E-H_{\mathrm{eff}}(\lambda)]=E^2-P_1(\lambda)E+P_2(\lambda),
\end{equation}
for which, the set of the only branch points becomes $S_{PQ}=  \{\lambda^{_{PQ}}_{1,l_1}, \lambda^{_{PQ}}_{2,l_2} \}$.

The branch-point set of $P_1(\lambda)$ and $P_2(\lambda)$ must be
the same as $S_{PQ}$\footnote{We can let $E=a$ and $b$ ($a,b$ are
	two unequal constants) respectively to construct two functions being
	analytic at the branch points of $E_1^{_P}(\lambda
	)$ and $E_2^{_P}(\lambda ) $ due to $PP$-crossing: $ f_1(\lambda)=a^2-[E_1^{_P}(\lambda
	)+E_2^{_P}(\lambda )]a+E_1^{_P}(\lambda )E_2^{_P}(\lambda )  $ and $
	f_2(\lambda)=b^2-[E_1^{_P}(\lambda )+E_2^{_P}(\lambda
	)]b+E_1^{_P}(\lambda )E_2^{_P}(\lambda )  $. Then $P_1(\lambda)=
	E_1^{_P}(\lambda )+E_2^{_P}(\lambda
	)=\frac{1}{b-a}[f_1(\lambda)-f_2(\lambda)]+b+a$ and $P_2(\lambda)=
	E_1^{_P}(\lambda )E_2^{_P}(\lambda
	)=\frac{1}{b-a}[(b-a+1)f_1(\lambda)-f_2(\lambda)]+ab$ certainly are
	also analytic at  these branch points. Such procedure can obviously
	be extended to the case of $N>2$.}.
Namely, unlike $
E^{_P}_1(\lambda) $ and $ E^{_P}_2(\lambda) $, $P_1(\lambda)$ and
$P_2(\lambda)$ also do not have branch points  due to the $PP$-crossing.

Therefore, unlike $ E^{_P}_1(\lambda) $- or $ E^{_P}_2(\lambda) $-series, the expansions
\begin{align*}
	P_1(\lambda)&=E_1^{_P}(\lambda )+E_2^{_P}(\lambda )\\
	&=\epsilon_1^{_P}+\epsilon_2^{_P}+\lambda (h^{\!_P}_{11}+h^{\!_P}_{22})
	+ \lambda^2\sum_{i=1,2}\sum_{n}\frac{|h^{\!_{P\!Q}}_{in}|^2}{\epsilon_i^{_P}-\epsilon_n^{_Q}}+\cdots,  \\
	P_2(\lambda)&=E_1^{_P}(\lambda )E_2^{_P}(\lambda ) =\epsilon_1^{_P}\epsilon_2^{_P}+\lambda \left(
	\epsilon^{_P}_1h^{_P}_{22}+\epsilon^{_P}_2h^{_P}_{11}\right) \\
	&\quad +\lambda^2\Big[
	h^{_P}_{11}h^{_P}_{22}- |h^{\!_P}_{12}|^2 + \sum_{n}\Big( \frac{\epsilon^{_P}_1|h^{\!_{P\!Q}}_{2n}|^2}{\epsilon_2^{_P}-\epsilon_n^{_Q}}
	+\frac{\epsilon^{_P}_2|h^{\!_{P\!Q}}_{1n}|^2}{\epsilon_1^{_P}-\epsilon_n^{_Q}} \Big) \Big]  +\cdots,
\end{align*}
only become divergent for $|\lambda|\geq r^{_P}$, where $r^{_P}=\min_{l_1,l_2}\{|\lambda^{_{PQ}}_{1,l_1}|, |\lambda^{_{PQ}}_{2,l_2}|\}$. The characteristic polynomial in Eq.~\eqref{2det} is further obtained and then the eigenvalues can be solved from $ \det[E-H_{\mathrm{eff}}(\lambda)]=0 $ as
\begin{equation}\label{eg2}
	E_{1,2}^{_P}(\lambda) =\frac{P_1(\lambda)\pm \sqrt{P_1(\lambda)^2-4P_2(\lambda)}}{2}.
\end{equation}

Such  procedure can be directly extended to a general case of $N$. Similar to the case of  $N=2$, we can use the $\lambda$-expansion of  $ E^{_P}_n(\lambda) $ in Eqs.~\eqref{EP1}--\eqref{EPN} to calculate $ P_{n}(\lambda) $ via Eqs.~\eqref{P1}-- \eqref{PN}. Also, since the branch-point set of  these $ P_{n}(\lambda) $ is the same as that of $\det[E-H_{\mathrm{eff}}(\lambda)]$,  the  convergence radius of the obtained  expansions can be generally determined as
\begin{equation}\label{rp}
	r^{_P}=\min_{l_1,l_2,\cdots,l_N}\{|\lambda^{_{PQ}}_{1,l_1}|, |\lambda^{_{PQ}}_{2,l_2}|,\cdots, |\lambda^{_{PQ}}_{N,l_N}| \}.
\end{equation}

Once the expansion of $ P_{n}(\lambda) $ has been obtained,  the detailed form of  $\det[E-H_{\mathrm{eff}}(\lambda)]$ can be calculated with Eq.~\eqref{edetN}. Then the $P$-space eigenvalues  $ E^{_P}_n(\lambda) $ can be solved just from
\begin{multline}\label{eqt}
	E^{N}-P_1(\lambda)E^{N-1}+\cdots+(-1)^{N-1}P_{N-1}(\lambda)E+(-1)^{N}P_{N}(\lambda)=0.
\end{multline}
Obviously, for $N<5$, Eq.~\eqref{eqt} can be solved analytically; while for $N\geq 5$, numerical methods can be employed.

Although such  procedure is still a perturbative one, we can
calculate $ E_{n}^{_P}(\lambda) $ for $|\lambda|<r^{_P}$ in
principle.
Comparing Eqs.~\eqref{rpb} and \eqref{rp},  we have $r^{_P}\geq \bar{r}^{_P}$. Actually, for cases that an effective Hamiltonian can be well-defined \cite{HS1}, the  $P$-  and $Q$-space eigenvalues are always ``well-separated'':
the $PQ$-crossing of eigenvalues only occurs when $|\lambda|$ becomes relatively large, compared with the $PP$-crossing, i.e.,
$r^{_P}> \bar{r}^{_P}$ or even $r^{_P}\gg \bar{r}^{_P}$; then, one can expect that with our procedure, $ E_{n}^{_P}(\lambda) $ can be calculated for a larger $\lambda$-region, even far beyond the applicable scope of a direct perturbation calculation.

Additionally, $\det[E-H_{\mathrm{eff}}(\lambda)]$ also can be
calculated directly with $H_{\mathrm{eff}}(\lambda)$.  Certainly,
this requires the  the specific form of $ H_{\mathrm{eff}}(\lambda)
$. However, due to the uniqueness of characteristic polynomial, any
expanded form of $ H_{\mathrm{eff}}(\lambda) $, regardless of the convergence radius,  can be used to calculate the coefficients  $ P_n(\lambda) $ and hence
$\det[E-H_{\mathrm{eff}}(\lambda)]$ uniquely in powers of $\lambda$.

\section{Simple example}

As a simple example, we consider a case of $N=2$ and $M=1$. Due to the small dimension, it is convenient to write $H(\lambda )$ in a matrix form, which we specify as
\begin{equation}\label{Hb}
	H(\lambda )=\begin{bmatrix}
		\epsilon_{1}^{_Q}  & 0 & 0\\
		0 & \epsilon_{2}^{_P} & 0\\
		0 & 0 & \epsilon_{1}^{_P}
	\end{bmatrix}+\lambda \begin{bmatrix}
		0  & h^{\!_{Q\!P}}_{12} & 0\\
		h^{\!_{P\!Q}}_{21} & 0 & h^{\!_P}_{21}\\
		0 & h^{\!_P}_{12} & 0
	\end{bmatrix},
\end{equation}
where we can take $ \epsilon_{1}^{_Q} =2$, $ \epsilon_{1}^{_Q} =1$, $ \epsilon_{2}^{_Q} =1.1$ and $ h^{\!_{P\!Q}}_{12} =h^{\!_{P\!Q}}_{21}=h^{\!_P}_{21}=h^{\!_P}_{12}=1 $ 
 (in arbitrary energy unit).

One purpose for us to value these quantities in this way is that in a perturbation expansion, the case of nearly-degenerate $\epsilon_{2}^{_P}$ and $\epsilon_{1}^{_P}$ is very important, since terms such as those containing ``$ \frac{1}{\epsilon_{2}^{_P}-\epsilon_{1}^{_P}}$'' would become very large and cause divergence. Such kind of divergence problem due to the quasi-degenerate unperturbed 
energies is well-known,  especially for the multireference perturbation
theory (MRPT) in studying multiconfigurational quantum-chemical systems, and is often called the ``intruder state problem'' (ISP) \cite{CM1,CM2,CM3,CM4,CM5}. As for our study of $ H_{\mathrm{eff}}(\lambda)
$, unperturbed 
energies $\epsilon_{n}^{_P}$ are always apart from $\epsilon_{m}^{_Q}$ far enough, and hence the ISP can occur mainly due to  the quasi-degenerate $\epsilon_{n}^{_P}$, or terms containing factors such as ``$
\frac{1}{\epsilon_{2}^{_P}-\epsilon_{1}^{_P}}$'', which we can call the ISP terms. 

This simple model can be analytically solved, and actually, a similar model has been extensively discussed in previous studies \cite{BP1,BP2}. To the order of $10^{-6}$, the  branch points due to the $PP$-crossing of $E_1^{_P}(\lambda )$ and $E_2^{_P}(\lambda )$ are $\lambda=\pm0.051392i$, and that due to  the $PQ$-crossing of $E_1^{_P}(\lambda )$ and $E_2^{_P}(\lambda )$ with $E_1^{_Q}(\lambda )$ are $\lambda=-0.238116\pm0.502871i$ and $0.238116\pm0.502871i$ respectively. Then, the convergence radius of the $E_1^{_P}(\lambda )$- and $E_2^{_P}(\lambda )$-series is $r_1^{_P}=r_2^{_P}\approx 0.051392$, while that for the valid calculation using the characteristic polynomial is $r^{_P}\approx 0.556397$.

\begin{figure}
	\centering
	\includegraphics[width=1\linewidth]{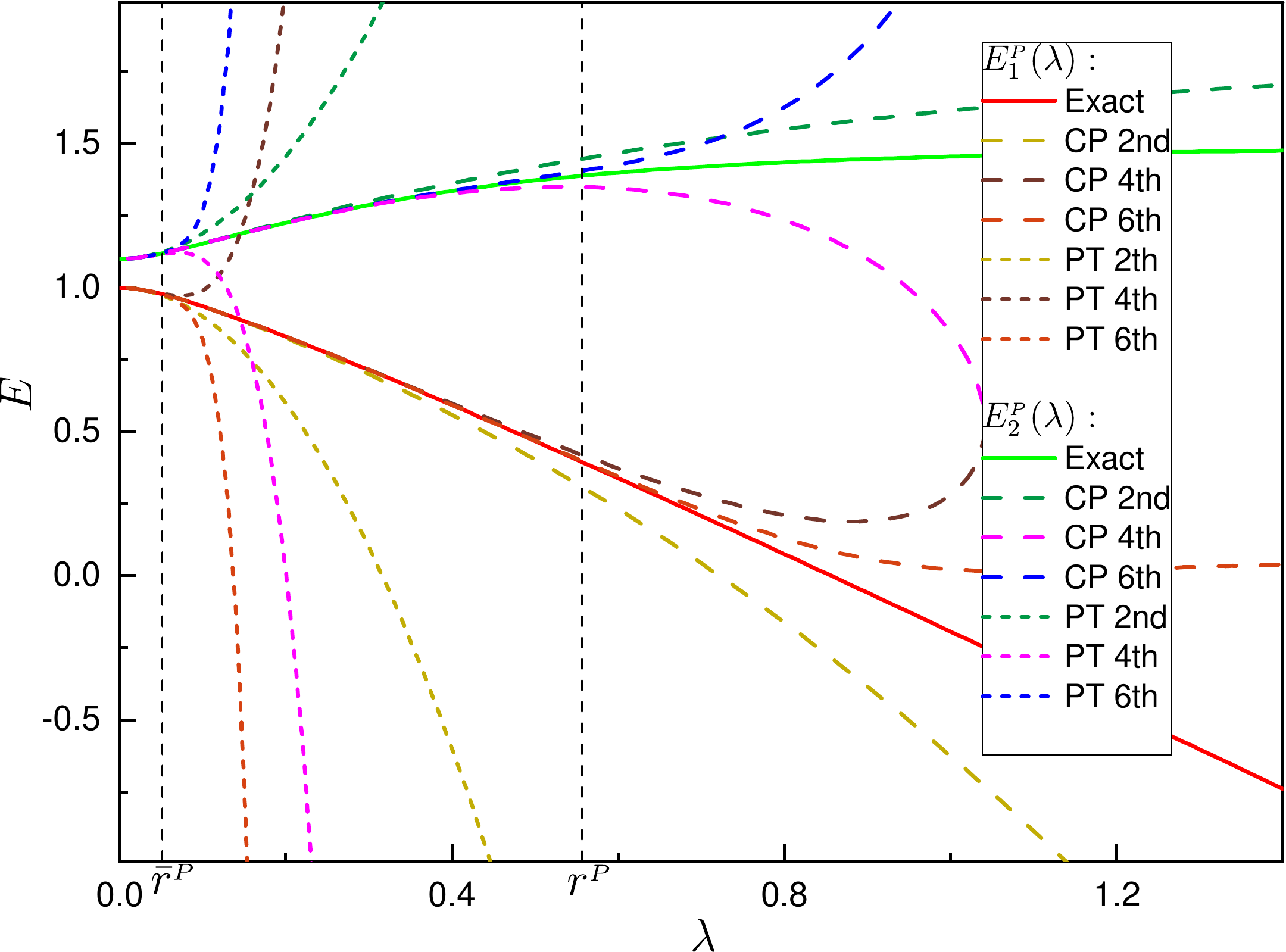}
	\caption{Results of $E_1^{_P}(\lambda )$ and $E_2^{_P}(\lambda )$ calculated respectively by characteristic polynomial (CP) for different-order (2nd, 4th and 6th) $\lambda$-expansion of coefficients, and by  different-order (2nd, 4th and 6th)  perturbation theory (PT), for the example model. Also shown is the exact value of $E_1^{_P}(\lambda )$ and $E_2^{_P}(\lambda )$. The position of $ \lambda=\bar{r}^{_P} $ and $ r^{_P} $ has been marked by vertical dash lines.}
	\label{fig1}
\end{figure}

We have calculated $E_1^{_P}(\lambda )$ and $E_2^{_P}(\lambda )$ via
our characteristic polynomial respectively for cases  in which only
terms through 2nd, 4th or 6th order are retained in the
$\lambda$-expansion of coefficients $ P_1(\lambda)$ and
$P_2(\lambda)$. The results are shown in Fig.~\ref{fig1}, and with
increasing order, the obvious deviation from the exact value of
$E_1^{_P}(\lambda )$ and $E_2^{_P}(\lambda )$ does tend to occur
at $\lambda=r^{_P}$. As a comparison, we have also shown the
result of $E_1^{_P}(\lambda )$ and $E_2^{_P}(\lambda )$  by the
usual perturbation theory, i.e., via retaining terms in
Eqs.~\eqref{EP1} and \eqref{EP2} through 2nd, 4th and 6th
respectively; with increasing order, the obvious deviation from the
exact value tends to occur at
$\lambda=r_1^{_P}=r_2^{_P}=\bar{r}^{_P}$, as expected.

An interesting thing should be noted is that in our calculation of
the characteristic polynomial, ISP terms always cancel in the coefficients $ P_1(\lambda)$
and $P_2(\lambda)$,  no matter for 2nd, 4th or 6th order
$\lambda$-expansion, leading to a final result being free from the
ISP.

Furthermore, as far as our simple example ($ N=2 $) is concerned, the branch points of $E_1^{_P}(\lambda )$ and $E_2^{_P}(\lambda )$ lying nearest to the origin can be alternatively determined by  $ P_1(\lambda)^2-4P_2(\lambda)=0 $, i.e., by the condition for $E_1^{_P}(\lambda )=E_2^{_P}(\lambda )$ from Eq.~\eqref{eg2}. When we retain terms in $ P_1(\lambda) $ and $ P_2(\lambda) $ to  the order of $\lambda^2$, $\lambda^4$, $\lambda^6$, such branch points are obtained as $\pm 0.0513922i$, $\pm 0.0513924i$, $\pm 0.0513922i$, respectively, of which the difference is vanish small. These values are  in good agreement with the exact one $\pm0.051392i$ obtained above.

\section{Discussion}

One advantage of using characteristic polynomial is its uniqueness.
There are infinitely many similar matrices that can be chosen as
$ H_{\mathrm{eff}}(\lambda) $; of which, the convergence radii for
$\lambda $-expansion obviously can be different, but all should not
exceed that of the characteristic-polynomial coefficients $
P_n(\lambda) $, i.e., $ r^{_P} $.  Mathematically, the most
convergent form of $ H_{\mathrm{eff}}(\lambda) $ can be obtained via
the least action of the unitary transformation to block diagonalize
$ H(\lambda) $, i.e., via changing $ H(\lambda) $ as little as
possible to bring it into a block diagonal form \cite{BL}. Such
``least action'' condition generally is hard to be satisfied. Hence,
one may expect that it is difficult to construct an  $
H_{\mathrm{eff}}(\lambda) $ which has the same convergence radius
for  $\lambda $-expansion as $ r^{_P} $. Actually, it has been shown
that the convergence radii of $ \lambda $-series expansion for those
$ H_{\mathrm{eff}}(\lambda) $ constructed by methods such as
Rayleigh-Schr\"{o}dinger or Brillouin-Wigner perturbation are mainly
determined by the  smallest module of the so-called
``exceptional points'' in the complex plane of $ \lambda  $ \cite{HS1, EP1},  where some  $P$-  and $Q$-state
eigenvalues, say $ E_1^{_P}(\lambda )$ and $ E_1^{_Q}(\lambda ) $,
coincide. These points should be distinguished from the branch
points, since they become the same only if the associated
eigenfunctions also coincide \cite{Ab1, EP1}. Obviously, each branch
point must also be an exceptional point, but the reverse is not
necessarily true. This means that for those $
H_{\mathrm{eff}}(\lambda) $, the convergence radius of $\lambda
$-expansion indeed cannot exceed $ r^{_P} $.

It is interesting whether there exists some unified way to construct $ H_{\mathrm{eff}}(\lambda) $ so that its  convergence radius for $\lambda $-expansion is always equal to $ r^{_P} $. The answer is yes. In fact, we can construct an effective Hamiltonian directly using the characteristic-polynomial coefficients $ P_n(\lambda) $ as follows
\begin{equation}\label{Heff}
	\bar{H}_{\mathrm{eff}}(\lambda)=\begin{bmatrix}
		\bar{P}_{1}(\lambda ) & \bar{P}_{2}(\lambda ) & \cdots & \bar{P}_{N-1}(\lambda ) & \bar{P}_{N}(\lambda ) \\
		1 & 0 & \cdots & 0 & 0 \\
		0 & 1 & \cdots & 0 & 0 \\
		\vdots &\vdots & \ddots& \vdots& \vdots \\
		0 & 0 & \cdots & 1 & 0
	\end{bmatrix},
\end{equation}
where $\bar{P}_{n}(\lambda )\equiv (-1)^{n+1}P_{n}(\lambda )$.
One can easily verify that the characteristic polynomial of $ \bar{H}_{\mathrm{eff}}(\lambda) $ is just in the same form shown in Eq.~\eqref{edetN} \cite{GZ1,GZ2}, and hence it indeed can be viewed an effective Hamiltonian. Obviously, due to its  relationship with $ P_n(\lambda) $, $ \bar{H}_{\mathrm{eff}}(\lambda) $ must have a convergence radius same as $ r^{_P} $ for $\lambda $-expansion.

Another point we want to discuss is whether our calculation of
$ P_n(\lambda) $ and $\det[E-H_{\mathrm{eff}}(\lambda)]$ can apply
to the case with degenerate or nearly-degenerate $ \epsilon_{n}^{_P}
$. In the simple example above, we have already mentioned that in
finite-order calculation of $ P_n(\lambda)$, ISP terms tend to
cancel. It is interesting whether this cancellation still
occurs when the order is very large or even tends to infinity, since as shown in Eqs~\eqref{EP1}--\eqref{EPN}, the $
E^{_P}_n(\lambda) $-series used to calculate $ P_{n}(\lambda) $
always have such terms .

We note that, due to the uniqueness of characteristic polynomial, if
one can construct any effective Hamiltonian $
H_{\mathrm{eff}}(\lambda) $ which can be expanded in powers of
$\lambda$ without the appearance of ISP terms, regardless of the
convergence radius, the complete cancellation of these terms in the
$\lambda $-expansion of $\det[E-H_{\mathrm{eff}}(\lambda)]$ or $
P_n(\lambda) $ can be proved.

Actually, the expanded form of $ H_{\mathrm{eff}}(\lambda) $
constructed in usual ways always does not have ISP terms, at least
when the first few orders of $\lambda$-expansion are retained
\cite{HS1, HS2, HS3, HS4, HS5, CT1,CT2}. A form of $
H_{\mathrm{eff}}(\lambda) $ which does not have any ISP terms for
all orders of $\lambda$-expansion is constructed in the Appendix.

Then, we find that ISP terms indeed are canceled completely in the
$\lambda $-expansion of $\det[E-H_{\mathrm{eff}}(\lambda)]$ or $
P_n(\lambda) $. This means that although non-degenerate unperturbed
energies have been assumed for $P$-space states, our discussion is
free from the ISP, and can directly apply to the cases with these
energies being nearly-degenerate or degenerate. Hence, our
discussion of the characteristic polynomial of
$H_{\mathrm{eff}}(\lambda)$ may find its potential use in MRPT in
quantum chemistry or other effective-Hamiltonian problems.

\section{Conclusion}
In conclusion, we have discussed the properties of the
characteristic polynomial of $H_{\mathrm{eff}}(\lambda)$ for a general model Hamiltonian. Unlike
$P$-space energy eigenvalues, $\det[E-H_{\mathrm{eff}}(\lambda)]$
and the coefficients $ P_n(\lambda) $ do not have the branch points
due to the $PP$-crossing, and hence are more suitable for a perturbation
calculation. We can calculate $ P_n(\lambda) $,  and hence $\det[E-H_{\mathrm{eff}}(\lambda)]$ in a perturbative way,
which can be further used to solve for the eigenvalues  $
E^{_P}_n(\lambda) $. Such procedure generally possesses a larger
convergence radius for $\lambda$ expansion, comparing with the
direct perturbation calculation of $ E^{_P}_n(\lambda) $. An
effective Hamiltonian $\bar{H}_{\mathrm{eff}}(\lambda)$, which has
the same branch points and convergence radius for $\lambda$
expansion as our characteristic polynomial,  has also been
constructed.

The
uniqueness of $\det[E-H_{\mathrm{eff}}(\lambda)]$ also brings
convenience to our discussion. We find that our procedure is free
from the ISP which arises when the unperturbed energies of $P$-space
states become nearly-degenerate or degenerate, as frequently
encountered in cases such as the MRPT study of multiconfigurational
quantum-chemical systems. Due to this advantage, one can expect that
our treatment of $\det[E-H_{\mathrm{eff}}(\lambda)]$ may find more
potential use in the future study of effective-Hamiltonian problems.

\appendix
\section{Constructing an effective Hamiltonian without the appearance of ISP terms}

To construct an
effective Hamiltonian without any ISP terms for our $H(\lambda)$, we
follow the  procedure in Refs.~\cite{HS2,Bh1,Bh2}. We can denote the
eigenstate associated with $E_n^{_P}(\lambda)$ by
$\left|\Psi_{n}\right\rangle$, i.e.,
\begin{equation*}
	H(\lambda)\left|\Psi_{n}\right\rangle=E_n^{_P}(\lambda)\left|\Psi_{n}\right\rangle, \quad n=1, \ldots, N.
\end{equation*}

We let $\hat{P}\equiv\sum_{n}|\psi^{_P}_n\rangle \langle \psi^{_P}_n|$ and
introduce
\begin{equation*}
	|\Psi^{_P}_n\rangle\equiv\hat{P}\left|\Psi_{n}\right\rangle,
\end{equation*}
where $ |\Psi^{_P}_n\rangle $  is required to be normalized, but $\left|\Psi_{n}\right\rangle$  is not. One can further introduce the so-called wave operator
$\hat{\Omega}$ by requiring that,
\begin{equation*}
	|\Psi_n\rangle=\hat{\Omega}\left|\Psi^{_P}_{n}\right\rangle.
\end{equation*}

It has been shown that $ \hat{\Omega}\hat{P}=\hat{\Omega}$ and $ \hat{P}\hat{\Omega}=\hat{P} $, from which we have
\begin{align}
	&\langle \psi^{_P}_m|\hat{\Omega} |\psi^{_P}_n\rangle=\langle \psi^{_P}_m|\hat{P}\hat{\Omega}|\psi^{_P}_n\rangle=\langle \psi^{_P}_m|\hat{P}|\psi^{_P}_n\rangle=\delta_{n,m}, \label{jy1}\\
	&\langle \psi^{_{P(Q)}}_m|\hat{\Omega} |\psi^{_Q}_n\rangle=\langle \psi^{_{P(Q)}}_m|\hat{\Omega} \hat{P} |\psi^{_Q}_n\rangle=0. \label{jy2}
\end{align}

An effective Hamiltonian can then be constructed as
\begin{multline} \label{eff}
	H_{\mathrm{eff}}(\lambda)=\hat{P} H_{0} \hat{P}+\lambda\hat{P} H_{I} \hat{\Omega} \\
	=\hat{P} H_{0} \hat{P}+\lambda\hat{P} H_{I}(\hat{\Omega}^{(0)}+\hat{\Omega}^{(1)}+\hat{\Omega}^{(2)}+\cdots)
\end{multline}
where $\hat{\Omega}$ has be  expanded in powers of  $ \lambda $  as $\hat{\Omega}=\hat{\Omega}^{(0)}+\hat{\Omega}^{(1)}+\hat{\Omega}^{(2)}+\cdots $,
with $ \Omega^{(k)} $  standing for the  $ k $th order term. Obviously, $\hat{\Omega}^{(0)}=\hat{P}$.

One can further derive an equation for $\hat{\Omega}$,
\begin{equation*}
	\left[\hat{\Omega}, H_{0}\right]=\lambda H_{I} \hat{\Omega}- \lambda \hat{\Omega}\hat{P}H_{I}\hat{\Omega},
\end{equation*}
with which an iterative formula for  $ \hat{\Omega}^{(k)} $  can be derived:
\begin{equation}\label{dd}
	\left[\hat{\Omega}^{(k)}, H_{0}\right]=\hat{A}_{k} \equiv \lambda
	H_{I}\hat{\Omega}^{(k-1)} -\lambda \sum_{j=0}^{k-1}
	\hat{\Omega}^{(j)}\hat{P}  H_{I} \hat{\Omega}^{(k-j-1)},
\end{equation}
where $ k=1,2, \cdots  $.

We can calculate $\hat{\Omega}^{(k)}$ in the bases of
$\{|\psi^{_P}_n\rangle, |\psi^{_Q}_m\rangle\}$. Due to
Eqs.~\eqref{jy1} and \eqref{jy2}, the nonzero matrix elements of
$\hat{\Omega}^{(k)}$ ($n\geq 1$) can only be those such as $\langle
\psi^{_{Q}}_m|\hat{\Omega}^{(k)} |\psi^{_P}_n\rangle$.

Now, we show that ISP terms, i.e., these
containing ``$ \frac{1}{\epsilon_{n_1}^{_P}-\epsilon_{n_2}^{_P}}$'', 
would not appear in the $ H_{\mathrm{eff}}(\lambda) $  constructed
above. Actually, all our procedure till now can be found in
Refs.~\cite{HS2,Bh1,Bh2}. Noting that $ \langle
\psi^{_{Q}}_m|[\hat{\Omega}^{(k)}, H_{0}]
|\psi^{_P}_n\rangle=(\epsilon_{n}^{_P}-\epsilon_{m}^{_Q})\langle
\psi^{_{Q}}_m|\hat{\Omega}^{(k)}|\psi^{_P}_n\rangle $, we have
\begin{equation*}
	\langle \psi^{_{Q}}_m|\hat{\Omega}^{(k)}|\psi^{_P}_n\rangle=\frac{\langle \psi^{_{Q}}_m|\hat{A}_{k} |\psi^{_P}_n\rangle}{\epsilon_{n}^{_P}-\epsilon_{m}^{_Q}},
\end{equation*}
from which, it follows that if $ \langle \psi^{_{Q}}_m|\hat{A}_{k}
|\psi^{_P}_n\rangle $, or equivalently, the matrix elements of
$\hat{\Omega}^{(j)}$, ($j=0,1,\cdots, k-1$), do not contain any ISP
terms, neither does $\langle
\psi^{_{Q}}_n|\hat{\Omega}^{(k)}|\psi^{_P}_m\rangle$.

Noting that ISP terms do not appear in $\hat{\Omega}^{(0)}=\hat{P}$,
we conclude that $\hat{\Omega}^{(1)}$, and hence
$\hat{\Omega}^{(2)},\hat{\Omega}^{(3)},\cdots$, all
would not have any ISP
terms appear in their matrix elements. Hence, when being expanded in
powers of  $ \lambda $, the $H_{\mathrm{eff}}(\lambda)$ given by
Eq.~\eqref{eff} indeed is an effective Hamiltonian without any ISP
terms.

\end{document}